\documentclass{article}%
\usepackage{amsmath}
\usepackage{amsfonts}
\usepackage{amssymb}
\usepackage{graphicx}%
\providecommand{\U}[1]{\protect\rule{.1in}{.1in}}
\begin{document}

\title{Entropic Dynamics and the Quantum Measurement Problem\thanks{Presented at
MaxEnt 2011, The 31st International Workshop on Bayesian Inference and Maximum
Entropy Methods in Science and Engineering, (July 10--15, 2011, Waterloo,
Canada). }}
\author{David T. Johnson and Ariel Caticha\\{\small Department of Physics, University at Albany-SUNY, }\\{\small Albany, NY 12222, USA.}}
\date{}
\maketitle

\begin{abstract}
We explore the measurement problem in the entropic dynamics approach to
quantum theory. The dual modes of quantum evolution---either continuous
unitary evolution or abrupt wave function collapse during measurement---are
unified by virtue of both being special instances of entropic updating of
probabilities. In entropic dynamics particles have definite but unknown
positions; their values are not created by the act of measurement. Other types
of observables are introduced as a convenient way to describe more complex
position measurements; they are not attributes of the particles but of the
probability distributions; their values are effectively created by the act of
measurement. We discuss the Born statistical rule for position, which is
trivially built into the formalism, and also for generic observables.

\end{abstract}

\section{Introduction}

Quantum mechanics introduced several new elements into physical theory. One is
indeterminism, another is the superposition principle embodied in both the
linearity of the Hilbert space and the linearity of the Schr\"{o}dinger
equation. Between them they dealt a very severe blow to the classical
conception of reality. The founders faced the double challenge of locating the
source of indeterminism and of explaining why straightforward consequences of
the superposition principle are not observed in the macroscopic world. Despite
enormous progress the challenge does not appear to have been met yet---at
least as evidenced by the number of questions that stubbornly refuse to go away.

The quantum measurement problem embodies most of these questions.\footnote{A
clear formulation of the problem is \cite{wigner:1963}; see also
\cite{ballentine:1998}. Modern reviews with references to the literature
appear in \cite{schlosshauer:2004} and \cite{jaeger:2009}.} One is the problem
of macroscopic entanglement; another is the problem of definite outcomes.
Since it is not possible to consistently assign objective values to physical
properties, when and how do values become actualized? How does a measurement
yield a definite outcome or how do events ever get to happen? Are the values
of observables created during the act of measurement?

An early \textquotedblleft solution\textquotedblright\ due to von Neumann
\cite{ballentine:1998} was to postulate a dual mode of wave function
evolution, either continuous and deterministic according to the
Schr\"{o}dinger equation, or discontinuous and stochastic during the
measurement process. It is in the latter process---the wave function collapse
or projection postulate \cite{komar:1962}\cite{ballentine:1990}---where
probabilities are introduced. Other proposed solutions involved denying that
collapse ever occurs which led to the many worlds, the many minds, and the
modal interpretations. These issues and others (such as the preferred basis
problem) can nowadays be tackled within the decoherence program
\cite{schlosshauer:2004} but with one strong caveat. Decoherence works but
only at the observational level---it saves the appearances. In this view
quantum mechanics is merely empirically adequate; it does not aim to provide
an objective picture of reality. Is this acceptable?

Our goal here is to revisit the problem of measurement from the fresh
perspective of Entropic Dynamics (ED) which introduces some new elements of
its own. \cite{caticha:2009}\cite{caticha:2011} In the standard view, which
remains popular to this day, quantum theory is considered an extension of
classical mechanics and therefore deviations from causality demand an
explanation. In the entropic view, on the other hand, quantum mechanics is an
example of entropic inference, a framework designed to handle insufficient
information. \cite{caticha:2008} From the entropic perspective indeterminism
requires no explanation. Uncertainty and probabilities are the norm; it is
certainty and determinism that demand explanations. The general attitude is
pragmatic \cite{stapp:1972}:\ physical theories are mere models for inference;
they do not attempt to mirror reality and, therefore, the best one can expect
is that they be empirically adequate, that is, good \textquotedblleft for all
practical purposes\textquotedblright. And this is not just the best one can
do, it is the best one ever needs to do. Therefore in the entropic framework
the program of decoherence is completely unobjectionable.

Once one accepts quantum theory as a theory of inference the dichotomy between
two distinct modes of wave function evolution is erased. Continuous unitary
evolution and discontinuous collapse correspond to two modes of processing
information, namely entropic updating in infinitesimal steps and Bayesian
updating in discrete finite steps. Indeed, as shown in \cite{caticha:2006}
these two updating rules are not intrinsically different; they are special
cases within a broader scheme of entropic inference.\cite{caticha:2008}

The other element that is significant for our present purpose is the
privileged role ascribed to the position observable. In ED, unlike the
standard interpretation of quantum mechanics, the positions of particles have
definite values just as they would in classical physics. Therefore the problem
of definite outcomes does not arise; the process of observation is essentially
classical. No inconsistencies arise because in ED position is the only
observable. More explicitly: other observables such as momentum, energy,
angular momentum and so on are not attributes of the particles but of the
probability distributions.\footnote{The case of momentum is discussed in
\cite{nawaz:2011}.} This opens the opportunity of explaining all other
\textquotedblleft observables\textquotedblright\ in purely informational terms.

After a brief review of background material on ED (section 2) we discuss the
measurement of observables other than position and derive the corresponding
Born rule (section 3). The issue of amplification is addressed in section 4
and we summarize our conclusions in section 5. A more detailed treatment of
the quantum measurement problem is given in \cite{johnson:2011}.

\section{Entropic Quantum Dynamics}

To set the context for the rest of the paper we brief{}ly review the three
main ideas that form the foundation of entropic dynamics. Several important
topics and most technical details are not discussed here. For a detailed
account of, for example, how time is introduced into an essentially atemporal
inference scheme, or the entropic nature of the phase of the wave function, or
the introduction of constants such as $\hbar$ or $m$, see \cite{caticha:2011}.
For simplicity here we discuss a single particle.

\noindent\textbf{The first idea} is about the subject matter: the goal is to
predict the position $x$ of a particle on the basis of some limited
information. We assume that in addition to the particle the world contains
other variables---we call them $y$. Not much needs to be known about the $y$
except that they are described by a probability distribution $p(y|x)$ that
depends on the particle position. The entropy of the $y$ variables is given
by
\begin{equation}
S[p,q]=-\int dy\,p(y|x)\log\frac{p(y|x)}{q(y)}=S(x)~.\label{entropy s}%
\end{equation}
Neither the underlying measure $q(y)$ nor the distribution $p(y|x)$ need to be
specified further. Note that $x$ enters as a parameter in $p(y|x)$ and
therefore its entropy is a function of $x$: $S[p,q]=S(x)$.

\noindent\textbf{The second idea} concerns the method of inference: we use the
method of maximum entropy subject to appropriate constraints to calculate the
probability $P(x^{\prime}|x)$ that the particle takes a short step from $x$ to
a nearby point $x^{\prime}$. The constraints reflect the relation between $x$
and $y$ given by $p(y|x)$, and the fact that motion happens gradually---a
large step is the result of many infinitesimally short steps. Thus
\emph{entropic dynamics does not assume any underlying sub-quantum mechanics
whether it be classical or not}.\footnote{And this is why the $y$ variables
are not hidden variables. The technical term `hidden variables' refers to
variables introduced to explain the emergent quantum behavior as a reflection
of an essentially classical dynamics -- whether stochastic or not -- operating
at a deeper level. The $y$ variables do not play this role because in ED there
is no underlying classical dynamics. } The successive accumulation of many
such short steps results in a probability distribution $\rho(x,t)$ that
satisfies the Fokker-Planck equation
\begin{equation}
\frac{\partial\rho}{\partial t}=-\vec{\nabla}\cdot\left(  \rho\vec{v}\right)
\label{FPEq}%
\end{equation}
where the current velocity $\vec{v}$ is
\begin{equation}
\vec{v}=\frac{\hbar}{m}\vec{\nabla}\phi\hspace{0.3in}\text{with}\hspace
{0.3in}\phi\left(  x,t\right)  =S\left(  x,t\right)  -\log\rho^{1/2}\left(
x,t\right)  ~.\label{v and phi}%
\end{equation}
These equations show how the entropy $S(x,t)$ guides the evolution of
$\rho(x,t)$.

\noindent\textbf{The third idea} is an energy constraint: the time evolution
of $S\left(  x,t\right)  $ is determined by imposing that a certain
\textquotedblleft energy\textquotedblright\ be conserved.\footnote{There is a
close parallel to statistical mechanics which also requires a clear
specification of the subject matter (the microstates), the inference method
(MaxEnt), and the constraints.} Thus, we require the diffusion to be
non-dissipative. To this end introduce an energy functional,%
\begin{equation}
E[\rho,S]=\int d^{3}x\rho\left(  x,t\right)  [\frac{\hbar^{2}}{2m}(\vec
{\nabla}\phi)^{2}+\frac{\hbar^{2}}{8m}(\vec{\nabla}\log\rho)^{2}+V~].
\end{equation}
Note that this energy is a statistical concept; it is not assigned to the
particle but to $\rho$ and $S$. Imposing that the energy be conserved for
arbitrary initial choices of $\rho$ and $S$ leads to the quantum
Hamilton-Jacobi equation,
\begin{equation}
\hbar\dot{\phi}+\frac{\hbar^{2}}{2m}(\vec{\nabla}\phi)^{2}+V-\frac{\hbar^{2}%
}{2m}\frac{\nabla^{2}\rho^{1/2}}{\rho^{1/2}}=0~.\label{QHJ}%
\end{equation}
This equation shows how the distribution $\rho(x,t)$ affects the evolution of
the entropy $S(x,t)$.

Finally, by combining the quantities $\rho$ and $S$ into a single complex
function, $\Psi=\rho^{1/2}e^{i\phi}$, the equations, (\ref{FPEq}) and
(\ref{QHJ}), can be rewritten into the Schr\"{o}dinger equation,%
\begin{equation}
i\hbar\frac{\partial\Psi}{\partial t}=-\frac{\hbar^{2}}{2m}\nabla^{2}%
\Psi+V\Psi~.\hspace{0.2in}\label{SE}%
\end{equation}
The fact that the Schr\"{o}dinger equation turned out to be linear and unitary
makes the language of Hilbert spaces and Dirac's bra-ket notation particularly
convenient---so from now we write $\Psi(x)=\langle x|\Psi\rangle$.

To conclude this brief review we emphasize that the Fokker-Planck equation
(\ref{FPEq}), the expression (\ref{v and phi}) for the current velocity as a
gradient, and the relation between phase $\phi$ and entropy $S$ are derived
and not postulated.

\section{Measurement in ED}

In practice the measurement of position can be technically challenging because
it requires the amplification of microscopic details to a macroscopically
observable scale. However, no intrinsically quantum effects need be involved:
the position of a particle has a definite, albeit unknown, value $x$ and its
probability distribution is, by construction, given by the Born rule,
$\rho(x)=|\Psi(x)|^{2}$. We can therefore assume that suitable position
detectors are available; in ED the measurement of position can be considered
as a primitive notion. This is not in any way different from the way
information in the form of data is handled in any other Bayesian inference
problem. The goal there is to make an inference on the basis of given data;
the issue of how the data was collected or itself inferred is not under
discussion. If we want, of course, we can address the issue of where the data
came from but this is a separate inference problem that requires an
independent analysis. In the next section we offer some additional remarks of
the amplification problem.

Our main concern here is observables other than position: how are they
defined, how are they measured? See \cite{johnson:2011} and
\cite{caticha:2000}. For simplicity, we will initially consider a measurement
that leads to a discrete set of possible position outcomes. In this case, the
continuous position probabilities become discrete,
\begin{equation}
\rho(x)\,dx=|\langle x|\Psi\rangle|^{2}\,dx\quad\rightarrow\quad
p_{i}=|\langle x_{i}|\Psi\rangle|^{2}\ .
\end{equation}

Since position is the only objectively real quantity there is no reason to
define other observables except that they may turn out to be convenient when
considering more complex experiments in which before the particles reach the
position detectors they are subjected to additional appropriately chosen
interactions, say magnetic fields or diffraction gratings. Suppose the
interactions within the complex measurement device $A$ are described by the
Schr\"{o}dinger eq.(\ref{SE}), that is, by a particular unitary evolution
$\hat{U}_{A}$. The particle will be detected at position $|x_{i}\rangle$ with
certainty provided it was initially in state $|a_{i}\rangle$ such that
\begin{equation}
\hat{U}_{A}|a_{i}\rangle=|x_{i}\rangle\ .
\end{equation}
Since the set $\{|x_{i}\rangle\}$ is orthonormal and complete, the
corresponding set $\{|a_{i}\rangle\}$ is also orthonormal and complete,
\begin{equation}
\langle a_{i}|a_{j}\rangle=\delta_{ij}\quad\text{and}\quad%
%TCIMACRO{\tsum \nolimits_{i}}%
%BeginExpansion
{\textstyle\sum\nolimits_{i}}
%EndExpansion
|a_{i}\rangle\langle a_{i}|{}=\hat{I}\ .\label{completeness}%
\end{equation}
Now consider the effect of this complex detector $A$ on some arbitrary initial
state vector $|\Psi\rangle$ which can always be expanded as
\begin{equation}
|\Psi\rangle=%
%TCIMACRO{\tsum \nolimits_{i}}%
%BeginExpansion
{\textstyle\sum\nolimits_{i}}
%EndExpansion
c_{i}|a_{i}\rangle\ ,
\end{equation}
where $c_{i}=\langle a_{i}|\Psi\rangle$ are complex coefficients. The state
$|\Psi\rangle$ will evolve according to $\hat{U}_{A}$ so that as it approaches
the position detectors the new state is
\begin{equation}
\hat{U}_{A}|\Psi\rangle=%
%TCIMACRO{\tsum \nolimits_{i}}%
%BeginExpansion
{\textstyle\sum\nolimits_{i}}
%EndExpansion
c_{i}\hat{U}_{A}|a_{i}\rangle=%
%TCIMACRO{\tsum \nolimits_{i}}%
%BeginExpansion
{\textstyle\sum\nolimits_{i}}
%EndExpansion
c_{i}|x_{i}\rangle\ .
\end{equation}
which, invoking the Born rule for position measurements, implies that the
probability of finding the particle at the position $x_{i}$ is
\begin{equation}
p_{i}=|c_{i}|^{2}\ .
\end{equation}

Thus, the probability that the particle in state $\hat{U}_{A}|\Psi\rangle$ is
found at position $x_{i}$ is $|c_{i}|^{2}$. But we can describe the same
outcome from the point of view of the more complex detector. The particle is
detected in state $|x_{i}\rangle$ as if it had earlier been in the state
$|a_{i}\rangle$. We adopt a new language and say, perhaps inappropriately,
that the particle has effectively been \textquotedblleft
detected\textquotedblright\ in the state $|a_{i}\rangle$, and therefore, the
probability that the particle in state $|\Psi\rangle$ is \textquotedblleft
detected\textquotedblright\ in state $|a_{i}\rangle$ is $|c_{i}|^{2}=|\langle
a_{i}|\Psi\rangle|^{2}$---which reproduces Born's rule for a generic
measurement device. The shift in language is not particularly fundamental---it
is a merely a matter of convenience but we can pursue it further and assert
that this complex detector \textquotedblleft measures\textquotedblright\ all
operators of the form $\hat{A}=$ $%
%TCIMACRO{\tsum \nolimits_{i}}%
%BeginExpansion
{\textstyle\sum\nolimits_{i}}
%EndExpansion
\lambda_{i}|a_{i}\rangle\langle a_{i}|$ where the eigenvalues $\lambda_{i}$
are arbitrary scalars. Born's rule is a postulate in the standard
interpretation of quantum mechanics; within ED we see that it is derived as
the natural consequence of unitary time evolution.

Note that it is not necessary that the operator $\hat{A}$ have real
eigenvalues, but it is necessary that its eigenvectors $|a_{i}\rangle$ be
orthogonal. This means that the Hermitian and anti-Hermitian parts of $\hat
{A}$ will be simultaneously diagonalizable. Thus, while $\hat{A}$ does not
have to be Hermitian ($\hat{A}=\hat{A}^{\dagger}$) it must certainly be
\emph{normal}, that is $\hat{A}\hat{A}^{\dagger}=\hat{A}^{\dagger}\hat{A}$.

Note also that if a sentence such as \textquotedblleft a particle has momentum
$\vec{p}$\textquotedblright\ is used only as a linguistic shortcut that
conveys information about the wave function before the particle enters the
complex detector then, strictly speaking, there is no such thing as the
momentum of the particle: the momentum is not an attribute of the particle but
rather it is a statistical attribute of the probability distribution $\rho(x)$
and entropy $S(x)$, a point that is more fully explored in \cite{nawaz:2011}.

The generalization to a continuous spectrum is straightforward. Let $\hat
{A}|a\rangle=a|a\rangle$. For simplicity we consider a discrete
one-dimensional lattice $a_{i}$ and $x_{i}$ and take the limit as the lattice
spacing $\Delta a=a_{i+1}-a_{i}\rightarrow0$. The discrete completeness
relation, eq. (\ref{completeness}),
\begin{equation}%
%TCIMACRO{\tsum \nolimits_{i}}%
%BeginExpansion
{\textstyle\sum\nolimits_{i}}
%EndExpansion
\Delta a\ \frac{|a_{i}\rangle}{(\Delta a)^{1/2}}\frac{\langle a_{i}|}{(\Delta
a)^{1/2}}=\hat{I}\quad\text{becomes}\quad\int\!da\ |a\rangle\langle
a|\,=\hat{I}\ ,
\end{equation}
where we defined%
\begin{equation}
\frac{|a_{i}\rangle}{(\Delta a)^{1/2}}\ \rightarrow\ |a\rangle\ .
\end{equation}

We again consider a measurement device that evolves eigenstates $|a\rangle$ of
$\hat{A}$ into unique position eigenstates $|x\rangle$, $\hat{U}_{A}%
|a\rangle=|x\rangle$. The mapping from $x$ to $a$ can be represented by an
appropriately smooth function $a=g(x)$. In the limit $\Delta x\rightarrow0$,
the orthogonality of position states is expressed by a Dirac delta
distribution,
\begin{equation}
\frac{\langle x_{i}|}{\Delta x^{1/2}}\frac{|x_{j}\rangle}{\Delta x^{1/2}%
}=\frac{\delta_{ij}}{\Delta x}\quad\rightarrow\quad\langle x|x^{\prime}%
\rangle=\delta(x-x^{\prime})\ .
\end{equation}
An arbitrary wave function can be expanded as
\begin{equation}
|\Psi\rangle=%
%TCIMACRO{\tsum \nolimits_{i}}%
%BeginExpansion
{\textstyle\sum\nolimits_{i}}
%EndExpansion
\Delta a\ \frac{|a_{i}\rangle}{\Delta a^{1/2}}\frac{\langle a_{i}|\Psi\rangle
}{\Delta a^{1/2}}\quad\text{or}\quad|\Psi\rangle=\int\!da\ |a\rangle\,\langle
a|\Psi\rangle\ .
\end{equation}
The unitary evolution $\hat{U}_{A}$ of the wave function leads to
\begin{align}
\hat{U}_{A}|\Psi\rangle & =%
%TCIMACRO{\tsum \nolimits_{i}}%
%BeginExpansion
{\textstyle\sum\nolimits_{i}}
%EndExpansion
\Delta a\ \frac{|x_{i}\rangle}{\Delta a^{1/2}}\frac{\langle a_{i}|\Psi\rangle
}{\Delta a^{1/2}}=%
%TCIMACRO{\tsum \nolimits_{i}}%
%BeginExpansion
{\textstyle\sum\nolimits_{i}}
%EndExpansion
\Delta x\ \frac{|x_{i}\rangle}{\Delta x^{1/2}}\frac{\langle a_{i}|\Psi\rangle
}{\Delta a^{1/2}}\left(  \frac{\Delta a}{\Delta x}\right)  ^{1/2}%
\ \nonumber\\
& \rightarrow\quad\int\!dx\ |x\rangle\,\langle a|\Psi\rangle|\frac{da}%
{dx}|^{1/2},
\end{align}
so that
\begin{equation}
p_{i}=|\langle x_{i}|\hat{U}_{A}|\Psi\rangle|^{2}=|\langle a_{i}|\Psi
\rangle|^{2}\quad\rightarrow\quad\rho(x)dx=|\langle a|\Psi\rangle|^{2}%
|\frac{da}{dx}|\,dx=\rho_{A}(a)da~.
\end{equation}
Thus, \textquotedblleft the probability that the particle in state $\hat
{U}_{A}|\Psi\rangle$ is found within the range $dx$ is $\rho(x)dx$%
\textquotedblright\ can be rephrased as \textquotedblleft the probability that
the particle in state $|\Psi\rangle$ is found within the range $da$ is
$\rho_{A}(a)da$\textquotedblright\ where
\begin{equation}
\rho_{A}(a)da=|\langle a|\Psi\rangle|^{2}\,da~,
\end{equation}
which is the continuum version of the Born rule for an arbitrary observable
$\hat{A}$.

\section{Amplification}

The technical problem of amplifying microscopic details so they can become
macroscopically observable is usually handled with a detection device set up
in an initial unstable equilibrium. The particle of interest activates the
amplifying system by inducing a cascade reaction that leaves the amplifier in
a definite macroscopic final state described by some pointer variable $\alpha$.

An eigenstate $|a_{i}\rangle$ evolves to a position $x_{i}$ and the goal of
the amplification process is to infer the value $x_{i}$ from the observed
value $\alpha_{r}$ of the pointer variable. The design of the device is deemed
successful when $x_{i}$ and $\alpha_{r}$ are suitably correlated and this
information is conveyed through a likelihood function $P(\alpha_{r}|x_{i}%
)$---an ideal amplification device would be described by $P(\alpha_{r}%
|x_{i})=\delta_{ri}$. Inferences about $x_{i}$ follow from a standard
application of Bayes rule,
\begin{equation}
P(x_{i}|\alpha_{r})=P(x_{i})\frac{P(\alpha_{r}|x_{i})}{P(\alpha_{r})}\ .
\end{equation}

The point of these considerations is to emphasize that there is nothing
intrinsically quantum mechanical about the amplification process. The issue is
one of appropriate selection of the information (in this case $\alpha_{r}$)
that happens to be relevant to a certain inference (in this case $x_{i}$).
This is, of course, a matter of design: a skilled experimentalist will design
the device so that no spurious correlations---whether quantum or
otherwise---nor any other kind of interfering noise will stand in the way of
inferring $x_{i}$.

It may seem that we are simply redrawing von Neumann's line between the
classical and the quantum with our treatment of the amplifying system. In some
sense, we are doing just that. However, the line here is not between a
classical \textquotedblleft reality\textquotedblright\ and a quantum
\textquotedblleft reality\textquotedblright---it is between the microscopic
particle with a definite but unknown position and an amplifying system
skillfully designed so its own microscopic degrees of freedom turn out to be
of no interest. In fact, in \cite{johnson:2011} we showed that such an
amplifier can be treated as a fully quantum system but it makes no difference
to the inference.

\section{Conclusions}

The solution of the problem of measurement within the entropic dynamics
framework hinges on two points: first, entropic quantum dynamics is a theory
of inference not a law of nature. This erases the dichotomy of dual modes of
evolution---continuous unitary evolution versus discrete wave function
collapse. The two modes of evolution turn out to correspond to two modes of
updating---continuous entropic and discrete Bayesian---which, within the
entropic inference framework, are unified into a single updating rule.

The second point is the privileged role of position---particles have definite
positions and therefore their values are not created but merely ascertained
during the act of measurement. All other \textquotedblleft
observables\textquotedblright\ are introduced as a matter of linguistic
convenience to describe more complex experiments. These observables turn out
to be attributes of the probability distributions and not of the particles;
their values are indeed \textquotedblleft created\textquotedblright\ during
the act of measurement.

\end{document}